\documentclass[twocolumn, english, longbibliography, superscriptaddress, breaklinks=true, showkeys, showpacs=false, nofootinbib]{revtex4}
\usepackage[T1]{fontenc}
\usepackage[latin9]{inputenc}
\usepackage{color}
\usepackage{babel}
\usepackage{amsmath}
\usepackage{amssymb}
\usepackage{subfigure}
\usepackage{graphicx}
\usepackage{physics}
\usepackage{comment}
\usepackage[unicode=true,pdfusetitle,
 bookmarks=true,bookmarksnumbered=false,bookmarksopen=false,breaklinks=true,pdfborder={0 0 0},backref=false,colorlinks=true]
 {hyperref}
\makeatletter
\@ifundefined{textcolor}{}
{%
 \definecolor{BLACK}{gray}{0}
 \definecolor{WHITE}{gray}{1}
 \definecolor{RED}{rgb}{1,0,0}
 \definecolor{GREEN}{rgb}{0,1,0}
 \definecolor{BLUE}{rgb}{0,0,1}
 \definecolor{CYAN}{cmyk}{1,0,0,0}
 \definecolor{MAGENTA}{cmyk}{0,1,0,0}
 \definecolor{YELLOW}{cmyk}{0,0,1,0}
}
\pdfoutput=1
\hypersetup{colorlinks=true,citecolor=blue,linkcolor=cyan,urlcolor=blue,filecolor= green, breaklinks=true}
\usepackage{url}
\usepackage{breakurl}
\makeatother

\begin{document}

\author{Diego S. Starke\href{https://orcid.org/0000-0002-6074-4488}{\includegraphics[scale=0.05]{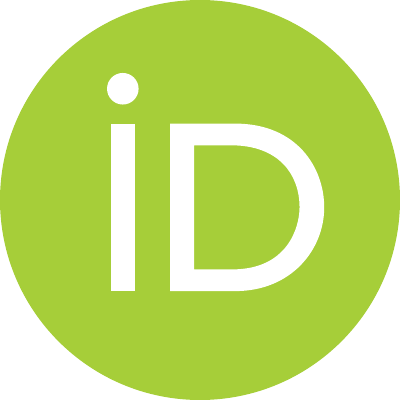}}}
\email{starkediego@gmail.com}
\affiliation{Physics Department, Center for Natural and Exact Sciences, Federal University of Santa Maria, Roraima Avenue 1000, Santa Maria, RS, 97105-900, Brazil}

\author{Marcos L. W. Basso\href{https://orcid.org/0000-0001-5456-7772}{\includegraphics[scale=0.05]{orcidid.pdf}}}
\email{marcoslwbasso@hotmail.com}
\affiliation{Center for Natural and Human Sciences, Federal University of ABC, Avenue of the States, Santo Andr\'e, S\~ao Paulo, 09210-580, Brazil}

\author{Jonas Maziero\href{https://orcid.org/0000-0002-2872-986X}{\includegraphics[scale=0.05]{orcidid.pdf}}}
\email{jonas.maziero@ufsm.br}
\affiliation{Physics Department, Center for Natural and Exact Sciences, Federal University of Santa Maria, Roraima Avenue 1000, Santa Maria, RS, 97105-900, Brazil}

\selectlanguage{english}%

\title{Efficient fidelity estimation: Alternative derivation and related applications}

\begin{abstract}
In [Phys. Rev. A 107, 012427 (2023)], A. J. Baldwin and J. A. Jones proved that Uhlmann-Jozsa's fidelity between two quantum states $\rho$ and $\sigma$, i.e., $F(\rho,\sigma)~:=~(Tr\sqrt{\sqrt{\rho}\sigma\sqrt{\rho}})^2$, can be written in a simplified form as $F(\rho,\sigma) = (Tr\sqrt{\rho\sigma})^2$. In this article, we give an alternative proof of this result, using a function power series expansion and the properties of the trace function. 
Our approach not only reinforces the validity of the simplified expression but also facilitates the exploration of novel dissimilarity functions for quantum states and more complex trace functions of density operators.
\end{abstract}

\keywords{Uhlmann-Jozsa fidelity, R\'enyi divergence, Efficient fidelity estimation, Chebyshev polynomials}

\date{\today}

\maketitle

The great promise of quantum computers is their ability to solve problems that are not possible within a reasonable time frame for classical computers, or even impossible for them~\cite{shor,nielsen,Harrow2017,Arute2019,Daley2022,Kim2023,Movassagh2023}. As our understanding of building quantum computers continues to advance, there is a growing desire for these systems to incorporate a greater number of qubits~\cite{1000,Manetsch2024,Zhang2023,Moses2023}. Furthermore, there is a recognition of the need for subsystems to have more than two levels, introducing the concept of qudits~\cite{goss,Wang2020,Chi2022,Deller2023,Kehrer2024}. This quest is driven by the promise of solving more complex problems, achieving greater computational efficiency, and unlocking new frontiers in quantum information processing. Whether through significant leaps in scale or the increase in dimensions, the expansion of quantum systems also introduces a greater complexity for the understanding of these systems. This inherent complexity in large systems provide a fertile ground for quantify performance, optimizing error correction, and exploring the unique advantages offered by the quantum realm. 

One way to assess the effectiveness of adjustments made in the quantum computers is by measuring the quantum fidelity between quantum states which are given by density operators. Quantum states describe the information that we have about the quantum system, once it is needed to calculate the probabilities of obtaining a specific measurement outcome of a given observable. Classically, one way to measure the similarity between the probabilities distributions $\mathbf{p}$ and $\mathbf{q}$ is using the classical fidelity, which is given by~\cite{matsumoto, nielsen}:
\begin{equation}
F(\mathbf{p},\mathbf{q}) = \left(\sum_{j=1}^d \sqrt{p_jq_j}\right)^2, \label{eq:fidc}
\end{equation}
where $\sum_{j=1}^d \sqrt{p_jq_j}$ is the Bhattacharyya coefficient. This expression quantifies how ``close'' two classical probability distributions are from each other.
It also provides a motivation in how to quantify the ``closeness'' of two quantum states by making appropriate modifications.

In Eq.~\eqref{eq:fidc}, the classical fidelity is introduced as a measure of similarity between probability distributions $\mathbf{p}$ and $\mathbf{q}$. Transitioning from classical to quantum settings involves the replacement of the classical probabilities $p_j$ and $q_j$ by the quantum mechanical probabilities $\Tr(\rho E_j)$ and $\Tr(\sigma E_j)$, respectively, where $\rho$ and $\sigma$ are density operators and $\{E_j\}$ are the elements of a positive operator-valued measure (POVM) satisfying $\sum_j E_j = \mathbb{I}$, with $\mathbb{I}$ being the identity operator \cite{Pinto2023}. It is worth mentioning that density operators can be interpreted as the quantum version of classical probability distributions. Therefore, in quantum mechanics, wherein the states are inherently more complex due to phenomena such as quantum superposition and entanglement, the natural extension for the fidelity $F(\rho,\sigma)$ between two density operators, as delineated in Ref.~\cite{fuchs}, is given by
\begin{equation}
F(\rho,\sigma) = \min_{\{E_j\}} \left(\sum_j \sqrt{\Tr(\rho E_j)}\sqrt{\Tr(\sigma E_j)}\right)^2, \label{eq:fid2}
\end{equation}
where the minimization is performed over all possible POVMs.

Analogous to the classical quantification of the similarity between probabilities distributions, in quantum mechanics the task is commonly accomplished through quantum fidelity, which can be used for the quantification of the proximity between a theoretically ideal quantum state and the state prepared by a quantum computer. However, it is interesting to notice that the first quantum fidelity measure for density operators was proposed by Jozsa~\cite{Jozsa} to describe the transition probability as introduced by Uhlmann~\cite{Uhlmann} and it is defined as
\begin{equation}
F(\rho,\sigma)~:=\left(\Tr\sqrt{\sqrt{\rho}\sigma\sqrt{\rho}}\right)^2. \label{eq:fido}
\end{equation}
This expression holds for any quantum states $\rho$ and $\sigma$.

The process to show that the definitions given by Eq.~\eqref{eq:fid2} and Eq.~\eqref{eq:fido} are equivalent can be summarized in the following steps: ($i$) the first step is to consider the unitary invariance an cyclic property of the trace function operation, i.e., $\Tr(\rho E_j) = \Tr(U\sqrt{\rho} E_j \sqrt{\rho} U^\dagger)$; ($ii$) the second step is the use of the Cauchy-Schwarz inequality, i.e, $\Tr(A^\dagger A)\Tr(B^\dagger B) \ge |\Tr(A^\dagger B)|^2$ for an appropriate choice of the operators $A$ and $B$, which implies that $\sqrt{\Tr(\rho E_j)}\sqrt{\Tr(\sigma E_j)} \ge \Tr(U \sqrt{\rho} E_j \sqrt{\sigma})$; ($iii$) the final step consists in summing over $j$ and choosing an appropriate form for the unitary operator $U$ in such way that the equality is attained, which is achieved by choosing $U = \sqrt{\sqrt{\rho}\sigma\sqrt{\rho}}\rho^{-1/2}\sigma^{-1/2}$. In Ref.~\cite{fuchs}, this demonstration is done rigorously and in more details.

Although there are various algorithms available for fidelity estimation \cite{wang,Cerezo2020,Leone2023}, the direct method for quantifying fidelity involves obtaining the quantum state through quantum state tomography (QST) \cite{qst,lundeen,tekkadath} to acquire the prepared state $\sigma$. Subsequently, matrix arithmetic operations are performed to compare the similarity between the prepared state $\sigma$ and the target theoretical state $\rho$. The higher the fidelity, the better the quantum computer can reproduce the desired theoretical state.
As we increase the number of subsystems of interest, the matrices involved in describing the quantum system grow exponentially. To calculate fidelity, we need to perform the diagonalization of a density matrix, calculate the product between them, and subsequently diagonalize the resulting matrix. Especially for very large density operators, obtaining diagonalization can be a demanding task since it is necessary to find the eigenvectors and eigenvalues numerically. Therefore, methods that approximate these values require careful consideration to ensure the desired precision.
The authors in Ref.~\cite{Baldwin} demonstrated a method to reduce the number of diagonalizations required for calculating the Uhlmann-Jozsa fidelity \cite{Uhlmann, Jozsa}, thereby improving computational efficiency. Their alternative form for fidelity relies on the fact that $\sqrt{\rho} \sigma \sqrt{\rho}$ is similar to $\rho \sigma$ and, therefore, shares the same set of eigenvalues, as noted in the work of Miszczak et al. \cite{miszczak}.

The quantification of fidelity remains one of the best ways to measure quantum similarity and assess the advancements achieved in quantum computers. Alternative methods include algorithms or integral forms utilizing matrix algebraic operations. However, the complexity of this quantification can pose a challenge, particularly with high-dimensional systems. Therefore, finding ways to simplify this process is desirable. In accordance to the Ref.~\cite{Baldwin}, we show in an alternative way the simplified expression obtained in their work. For that, our alternative proof relies on power series expansion and the properties of the trace function. By solely leveraging the properties of the trace function and assuming it is feasible to express a function composed of three linear operators in a series of polynomials, we will demonstrate that $Tr\{f(ABC)\} = Tr\{f(CAB)\} = Tr\{f(BCA)\}$.

Let us consider three linear operators $A,B,C$ defined on a Hilbert space $\mathcal{H}$. We begin by writing 
a function of the product of these three linear operators that can be expanded as a power series:
\begin{equation}
f(ABC) = \sum_{j=0}^\infty c_j (ABC)^j. \label{eq:pseries}
\end{equation}
Using the cyclic property of the trace function, we obtain
\begin{align}
\Tr\left\{(ABC)^j\right\} & = \Tr(ABCABC\cdots ABC) \nonumber\\
& = \Tr(CABCAB\cdots CAB) \nonumber\\
& = \Tr\left\{(CAB)^j\right\}. 
\end{align}
With this, using the linearity of the trace function, we can write
\begin{align}
\Tr\{f(ABC)\} & = \Tr\left\{\sum_{j=0}^\infty c_j (ABC)^j\right\} \nonumber\\
& = \sum_{j=0}^\infty c_j \Tr\{(ABC)^j\} \nonumber\\
& = \sum_{j=0}^\infty c_j \Tr\{(CAB)^j\} \nonumber\\
& = \Tr\left\{\sum_{j=0}^\infty c_j (CAB)^j\right\} \nonumber\\
& = \Tr\{f(CAB)\}. \label{eq:trfabc}
\end{align}

As a first application of this general result, we regard the Uhlmann-Jozsa fidelity between two density operators $\rho$ and $\sigma$, that is defined as in Eq. (\ref{eq:fido}).
So, by applying Eq.~(\ref{eq:trfabc}) with 
\begin{equation}
f(x)=\sqrt{x},\ A=C=\sqrt{\rho},\ B=\sigma,
\end{equation}
we will have
\begin{align}
\!\!\!F(\rho,\sigma) & = \left(\Tr\sqrt{\sqrt{\rho}\sigma\sqrt{\rho}}\right)^2  = \left(\Tr\sqrt{\sqrt{\rho}\sqrt{\rho}\sigma}\right)^2 \nonumber\\ 
& = \left(\Tr\sqrt{\rho\sigma}\right)^2, \label{eq:fid}
\end{align}
which is the result obtained in Ref.~\cite{Baldwin}. 

If the power series considered in Eq.~\eqref{eq:pseries} is the Taylor series, then this alternative derivation of Eq.~\eqref{eq:fid} is valid only for positive-definite hermitian density operators, once the derivatives of the square root function are not defined for null eigenvalues. However, this derivation can be extended for positive semi-definite hermitian density operators by using the Chebyshev polynomials~\cite{mason2003,weisse2008}, as done similarly in Ref.~\cite{Wihler} for the von-Neumann entropy. The Chebyshev polynomials are presented in Appendix~\ref{app:Chebyshev}, where we also describe the expansion of an arbitrary function in terms of the Chebyshev polynomials as a power series in order for the derivations of Eqs.~\eqref{eq:trfabc} and~\eqref{eq:fid} to remain valid. Moreover, since  $\sqrt{\rho} \sigma \sqrt{\rho}$ is hermitian and also positive semi-definite, its eigenvalues are real and are defined in the interval $[0,b]$, where $b \in \mathbb{R}$. This interval can be mapped in the Chebyshev interval as shown in Appendix~\ref{app:Chebyshevsub1}.

Now, let us consider another application of our result expressed by Eq.~\eqref{eq:trfabc} regarding alternative quantum states dissimilarity functions. For instance, the $\alpha$-$z$-relative R\'enyi entropy \cite{audenaert,Lin2015,Androulakis2024} is defined as:
\begin{equation}
\!\!\!D_{\alpha,z}(\rho||\sigma) \!=\! \frac{1}{\alpha-1}\log \Tr\big[\sigma^{(1-\alpha)/2z}\rho^{\alpha/z}\sigma^{(1-\alpha)/2z}\big]^z\!\!,
\end{equation}
with $\alpha\in\mathbb{R}$ and $z \in \mathbb{R}^+$. Let us notice that $D_{\alpha,1}(\rho||\sigma)=D_{\alpha}(\rho||\sigma)$ and $D_{\alpha,\alpha}(\rho||\sigma)=\tilde{D}_{\alpha}(\rho||\sigma),$ i.e., $D_{\alpha,z}(\rho||\sigma)$ generalizes the R\'enyi divergence of order $\alpha$, $D_\alpha (\rho||\sigma) = \frac{1}{\alpha -1}\log \Tr(\rho^\alpha \sigma^{1-\alpha})$,
and the sandwiched R\'enyi divergence of order $\alpha$, 
$\tilde{D}_\alpha (\rho||\sigma) = 
\frac{1}{\alpha-1}\log \Tr\big[\sigma^{(1-\alpha)/2\alpha}\rho\sigma^{(1-\alpha)/2\alpha}\big]^\alpha$. Besides, the relative entropy is obtained as the following limit: $\lim_{\alpha\rightarrow 1} D_\alpha (\rho||\sigma) = D(\rho||\sigma) = \Tr(\rho(\log\rho-\log\sigma)).$ 
Now, applying our result $\Tr(f(ABC)) = \Tr(f(CAB))$
with
\begin{equation}
f(x)=x^z,\ A=C=\sigma^{(1-\alpha)/2z},\ B = \rho^{\alpha/z},
\end{equation}
we can write
\begin{align}
\!\!\!\!D_{\alpha,z}(\rho||\sigma) &\! = \!\frac{1}{\alpha-1}\log \Tr\big[\sigma^{(1-\alpha)/2z}\rho^{\alpha/z}\sigma^{(1-\alpha)/2z}\big]^z \nonumber\\
&\! =\! \frac{1}{\alpha-1}\log \Tr\big[\sigma^{(1-\alpha)/2z}\sigma^{(1-\alpha)/2z}\rho^{\alpha/z}\big]^z \nonumber\\
& \!=\! \frac{1}{\alpha-1}\log \Tr\big[\sigma^{(1-\alpha)/z}\rho^{\alpha/z}\big]^z .
\end{align}
Let us notice that we did not avoid any matrix diagonalization, however we need one less matrix multiplication calculation for obtaining $D_{\alpha,z}(\rho||\sigma)$.

On the other hand, regarding
\begin{align}
D_{\alpha,\alpha}(\rho||\sigma) & = \frac{1}{\alpha-1}\log \Tr\big[\sigma^{(1-\alpha)/\alpha}\rho^{\alpha/\alpha}\big]^\alpha \nonumber\\
& = \frac{1}{\alpha-1}\log \Tr\big[\sigma^{(1-\alpha)/\alpha}\rho\big]^\alpha,
\end{align}
we can avoid one matrix diagonalization in the following particular case
\begin{align}
D_{\frac{1}{2},\frac{1}{2}}(\rho||\sigma) & = \frac{1}{1/2-1}\log \Tr\big[\sigma^{(1-1/2)/(1/2)}\rho\big]^{1/2} \nonumber\\
& = -2\log \Tr\sqrt{\sigma\rho} \nonumber\\
& = -2\log(F(\rho,\sigma)).
\end{align}
We observe that this equality was already reported in Ref. \cite{audenaert}.

To sum up, in this article we gave an alternative derivation of the simplified expression obtained by Ref.~\cite{Baldwin} for the Uhlmann-Jozsa fidelity. Our result relies solely on the expansion in power series of a function of the product of three linear operators and the cyclic and linearity properties of the trace function.  Moreover, we also obtained an alternative expression for others quantum states dissimilarity functions. Finally, we hope that our method will enable further simplifications of other expressions that may arise in the study and development of quantum information science.

Finally, we observe that after the completion of this manuscript, we became aware of the related work reported in Ref.~\cite{muller}.

\begin{acknowledgments}
This work was supported by the Coordination for the Improvement of Higher Education Personnel (CAPES), Grant No. 88887.827989/2023-00, by the 
S\~ao Paulo Research Foundation (FAPESP), Grant No.~2022/09496-8, by the National Council for Scientific and Technological Development (CNPq), Grants No. 309862/2021-3, No. 409673/2022-6, and No. 421792/2022-1, and by the National Institute for the Science and Technology of Quantum Information (INCT-IQ), Grant No. 465469/2014-0.
\end{acknowledgments}

\vspace*{1cm}

\appendix

\section{Power Series by Chebyshev Polynomials}
\label{app:Chebyshev}
Let us first remember that any square-integrable and continuous function $f(x)$ in a given interval can be expanded in terms of a complete set of orthogonal functions~\cite{byron}. Hence, the Chebyshev polynomials~\cite{mason2003} can be used to expand a given function $f(x)$, i.e.,
\begin{equation}
f(x) = \frac{1}{2}a_0T_0(x) + \sum_{n=1}^{\infty} a_n T_n(x),
\end{equation}
where $T_n(x)$ denotes the Chebyshev polynomials of the first kind and can be defined through the recurrence relation
\begin{equation}
T_{n+1}(x) = 2xT_n(x) - T_{n-1}(x), \quad n\ge 1,
\end{equation} 
with $T_0(x) = 1$ and $T_1(x) = x$. We notice that there are other ways to define the Chebyshev polynomials~\cite{mason2003}. However, the definition above is sufficient for our purposes here.

The Chebyshev polynomials obey the orthogonality relation expressed as
\begin{equation}
\int_{-1}^{1}  T_{n}(x) T_{m}(x)w(x)dx =
\begin{cases}
0, & \text{if } n \neq m,\\
\pi, & \text{if } n = m = 0,\\
\frac{\pi}{2}, & \text{if } n = m \neq 0,
\end{cases}
\end{equation}
where $w(x) = 1/\sqrt{1 - x^2}$ is the weight function.

The expansion coefficients are given by
\begin{equation}
a_n = \frac{2}{\pi} \int_{-1}^{1} f(x) T_n(x)w(x) dx, \quad n \geq 0.
\end{equation}
Here $f(x)$ is the desired mapped function to the Chebyshev interval $[-1,1]$. For instance, if the interval of $f(x)$ is $[a,b]$, then the transformation $\phi(x) = 2(x-a)/(b-a) - 1$ maps the interval $[a,b]$ to the Chebyshev interval $[-1,1]$.

Moreover, we can write the Chebyshev polynomials in a power series as follows
\begin{equation}
T_n(x) = \sum_{j=0}^n a_{n,j} x^j,
\end{equation}
which implies that the function $f(x)$ can be expressed as
\begin{equation}
f(x) = \sum_{j=0}^{n} \sum_{n = 0}^{\infty} \eta_{n,j}x^j, \label{eq:f(x)}
\end{equation}
where $\eta_{0,0} = a_0/2$ and $\eta_{n,j} = a_na_{n,j}$ for $n\ge 1$. Eq.~\eqref{eq:f(x)} allows us to follow the same path as outlined in the main text and obtain the Eq.~\eqref{eq:trfabc}, which is needed in order to derive the alternative expression for the Uhlmann-Jozsa fidelity and the $\alpha$-$z$-relative R\'enyi entropy.

The last step for deriving the alternative expression for the Ulhman-Josza fidelity and the $\alpha$-$z$-relative R\'enyi entropy is the mapping between the interval $[0,b]$ and the Chebyshev interval $[-1,1]$ for an arbitrary function. To do this, let us notice that, for $b > 0$, the following transformation 
\begin{align}
    &\phi: [-1,1] \to [0,b], \nonumber \\ 
    &\phi(x) = \frac{b}{2}(x+1),  \\
    &\phi^{-1}(x) = \frac{2}{b}x -1, \nonumber
\end{align}
allow us to define the Chebyshev polynomials $\mathcal{T}_n$ on the interval $[0,b]$, where $\mathcal{T}_n = T_n \circ \phi^{-1}$.

Finally, for the Ulhman-Josza fidelity, the function to be expanded is $f(x) = \sqrt{x}$, which is continuous and square-integrable in the Chebyshev interval. While, for the  $\alpha$-$z$-relative R\'enyi entropy, the function to be expanded is $f(x) = x^z$ for $z \in \mathbb{R}^+$, which is also continuous and square-integrable in the Chebyshev interval. Therefore, we can safely apply the expansion through the Chebyshev polynomials. Therefore, to show that this expansion works, we choose to present below the explicit calculations only for $f(x) = \sqrt{x}$, due to its simplicity.

\subsection{Uhlmann-Jozsa fidelity function}
\label{app:Chebyshevsub1}

Consider that $f(x) = \sqrt{x}$ in  the interval $[0,b]$. The new function mapped in the Chebyshev interval $[-1,1]$ is given by
\begin{equation}
f(x) = \sqrt{\frac{b(x+1)}{2}}.
\end{equation}

The coefficients are obtained by
\begin{align}
a_n &= \frac{\sqrt{2b}}{\pi} \int_{-1}^{1} \sqrt{\frac{x+1}{1 - x^2}} T_n(x) dx \nonumber\\
&= \frac{\sqrt{2b}}{\pi} \int_{-1}^{1} \frac{T_n(x)}{\sqrt{1 - x}} dx,
\end{align}
which has a nice closed form given by
\begin{equation}
a_n =\frac{4\sqrt{b}}{\pi}\frac{(-1)^{n+1}}{4n^2-1}.
\end{equation}
This solution is derived through induction and involves the calculation of well known integrals of the type~\cite{schaum}
\begin{align}
\int \frac{x^kdx}{\sqrt{\gamma x + \beta}} = \frac{2x^k\sqrt{\gamma x + \beta}}{(2k + 1)\gamma }\nonumber\\
 - \frac{2k\beta}{(2k + 1)\gamma } \int \frac{x^{k-1}}{\sqrt{\gamma x + \beta}} \, dx,
\end{align}
which, for our case, $k\in \mathbb{N}$, $\gamma = -1$ and $\beta = 1$.


\end{document}